\documentclass[10pt,journal,letterpaper]{IEEEtran}

\IEEEoverridecommandlockouts
\usepackage[T1]{fontenc}
\usepackage{orcidlink}

\usepackage{setspace}

\IEEEoverridecommandlockouts
\usepackage{fancyhdr}
\usepackage[numbers, sort&compress]{natbib}
\usepackage{url}
\usepackage{graphicx}

\usepackage{float}
\usepackage{enumerate}
\usepackage{hyperref}
\usepackage{amsmath,amssymb,amsfonts}
\usepackage{algorithmic}
\usepackage[ruled,vlined]{algorithm2e}
\usepackage{graphicx}
\usepackage{array}     
\usepackage{amsmath}   
\usepackage{textcomp}
\usepackage{xcolor}
\usepackage{physics}
\usepackage[caption=false]{subfig}
\usepackage[font=footnotesize, skip=0pt,compatibility=false]{caption}
\usepackage{subcaption}
\usepackage{comment}
\usepackage{tikz}
\usetikzlibrary{positioning, shapes, arrows.meta}
\usepackage{booktabs}
\usepackage{multirow}
\def\BibTeX{{\rm B\kern-.05em{\sc i\kern-.025em b}\kern-.08em
    T\kern-.1667em\lower.7ex\hbox{E}\kern-.125emX}}
\begin{document} 
\bstctlcite{MyBSTcontrol}
\title{A New Quantum Secure Time Transfer  System}

\author{Ravi Singh Adhikari$^{*}$, Aman Gupta$^{*}$, Anju Rani$^{*}$, Xiaoyu~Ai$^{*}$, and~Robert~Malaney$^*$ 
\thanks{$^*$Ravi Singh Adhikari, Aman Gupta, Anju Rani,   Xiaoyu Ai, and Robert Malaney are with the School of Electrical Engineering and Telecommunications, University of New South Wales, Sydney, Australia.}
}

\maketitle
\thispagestyle{empty}
\pagestyle{empty}
\begin{abstract}
High-precision clock synchronization is essential for a wide range of network-distributed applications. In the quantum space, these applications include communication, sensing, and positioning. However, current synchronization techniques are vulnerable to attacks, such as intercept-resend attacks, spoofing, and delay attacks. Here, we propose and experimentally demonstrate a new quantum secure time transfer (QSTT) system, subsequently used for clock synchronization, that largely negates such attacks. Novel to our system is the optimal use of self-generated quantum keys within the QSTT to information-theoretically secure the maximum amount of timing data; as well as the introduction, within a hybrid quantum/post-quantum architecture, of an information-theoretic secure obfuscated encryption sequence of the remaining timing data. With these enhancements, we argue that our new system represents the most robust implementation of QSTT to date.

\end{abstract}

\section{Introduction}\label{sec:introduction}

Precise time-synchronized remote clocks are essential for a wide range of modern-day applications, including positioning, navigation,  and telecommunications. In the quantum space, precise synchronization is also critical for applications such as
distributed quantum computing, distributed quantum sensing, and quantum communication~\cite{doi:10.1126/science.aam9288}. 

Classical clock synchronization methods such as the Global Positioning System~(GPS) time transfer~\cite{1537364}, the Network Time Protocol~\cite{103043}, and the White Rabbit Protocol~\cite{6070148} are widely used for time services. However, these approaches suffer from limited timing precision and vulnerability to attacks such as spoofing, jamming, delay attacks, and other cyber-attacks~\cite{alghamdi2020cyber,alghamdi2021precision}. 
To address the limitations of classical clock synchronization, various quantum clock synchronization techniques and time transfer protocols have been proposed and demonstrated~\cite{PhysRevLett.85.2010, Giovannetti2001, 10.1063/1.5086493, Quan:22, 10.1063/1.5121489, PhysRevApplied.22.024012,PhysRevA.100.023849,lafler2023quantumtimetransferpractical}. A quantum technique widely used for clock synchronization is that which exploits the tight temporal correlation properties of energy-time entangled photon pairs generated through a quantum process called spontaneous parametric down-conversion (SPDC)~\cite{Jabir2017, PhysRevApplied.19.054082}. In this technique, the entangled photons serve as timing signals for time transfer, with timing information encoded in their arrival times. Using entangled photon pairs for time transfer, femtosecond-level precision in clock synchronization has been demonstrated~\cite{quan2018experimental}.   

These entanglement-based protocols remain vulnerable to man-in-the-middle attacks because the timing  information (sent in a classical channel) is neither authenticated nor encrypted. Using this timing information, the attacker can unsynchronize two clocks or synchronize their own clock to attack other time-based systems that rely on precise clock synchronization~\cite{motero2021attacking, breitinger2025sok,mastromauro2025survey}. To address these vulnerabilities, encryption and authentication are required. To this end, here we introduce a new quantum secure time transfer (QSTT) protocol embedded within a combined quantum plus post-quantum obfuscated architecture (we refer to the combined QSTT protocol plus obfuscated architecture as our QSTT system). Our obfuscated architecture, which combines quantum key distribution (QKD) and post-quantum cryptography (PQC) is described in detail elsewhere~\cite{rani2025obfuscatedquantumpostquantumcryptography}. The specific elements of that architecture we apply to our QSTT system will be detailed later.

We caution that across the existing literature the phrase QSTT has been taken to mean slightly different aspects of quantum enhancement, this enhancement is usually applicable under different assumptions (\textit{e.g}.,~\cite{Dai2020, villas}). The uninitiated reader may be convinced that QSTT means that all data associated with the time transfer  are information-theoretic secured via a QKD one-time pad (OTP). However, in practice, this is usually impossible - the creation rate of timing data in deployed QKD protocols almost certainly is greater than the secret key creation rate from QKD. For this reason, in~\cite{Dai2020} the encryption of the timing data is via the Advanced Encryption Standard (AES), with QKD utilized only as the AES seed. However, this leads to a situation where QSTT has no embedded information-theoretic security; AES is a method that only provides unproven security against known attacks, including those based on quantum computer algorithms. Improving on this unsatisfactory situation is the main focus of this work.

To enable progress, we provide the following new definition.
We define QSTT  as a protocol in which quantum signaling is used for time transfer and secure key generation, \textit{ where QKD generated keys are used to encrypt in an information-theoretic manner the maximum amount of timing data restricted on $r_1\le r_2$, where $r_1$  denotes the  QKD key usage rate and $r_2$  the QKD key creation rate}. 
 Our QSTT system should be considered separate from the synchronization protocol. The former is used to secure the data needed for synchronization, whereas the latter describes the synchronization procedure that uses those data.

The contributions of this work are summarized as follows. (i) We design a new QSTT system that is used to provide secure network clock synchronization. An ingredient of this QSTT is the QKD-OTP encryption of the maximum amount of timing data constrained on  $r_1\le r_2$, thereby optimizing the use of limited QKD resources.
    (ii)  The timing data \textit{not} QKD-OTP encrypted due to these limited resources is instead encrypted via a sequence of post-quantum solutions, this sequence being information-theoretic secured. In addition, a scrambling of the temporal sequence of all the timing data is invoked.
    (iii) An implementation of the proposed QSTT system and the synchronization to which it leads is presented.

\section{System Model}\label{sec: Synchronization protocol}

In our previous work~\cite{rani2025obfuscatedquantumpostquantumcryptography}, we demonstrated a QKD-PQC implementation for data messages.
An obfuscated sequence determined the order in which the encryption-decryption algorithms (QKD and multiple PQC solutions) were applied to the message. This sequence was then encrypted using a pre-shared key (PSK) known only to the sender-receiver pair, ensuring information-theoretic secure transmission of the sequence (the PSK is the \textit{a priori} key required by QKD).
By this means, the encryption process was unknown to an adversary, providing an additional layer of practical security for the encrypted data message. A synchronization process was invoked in~\cite{rani2025obfuscatedquantumpostquantumcryptography}, however, this did not involve QSTT in any form. Instead, it invoked
a Global Positioning System (GPS) free quantum synchronization protocol 
that did not incorporate an obfuscated sequence for the timing data or the optimal use of QKD-OTP encryption. We explore and implement these new features here. For clarity of exposition, in the following we refer to the terms `QKD keys' and `PQC keys' to mean keys that are generated via the QKD and PQC key encapsulation protocols, respectively. Similarly, we refer to the terms `QKD-OTP' and `PSK-OTP' to mean one-time padding (XoR operation) with QKD and pre-shared keys, respectively. Here, we assume that PSK-OTP always provides information-theoretic security and that  QKD-OTP provides the same only under the assumption that no side-channel attack on QKD has been successfully executed.

 In the experimental implementation we pursue, the quantum signals are entangled photon pairs generated via the SPDC process.  An entangled-photon source has become a prominent candidate for synchronization processes because the photon pairs are highly temporally correlated with each other, typically within the femtosecond regime~\cite{maclean2018direct}. 
 The complete theoretical analysis of the SPDC process for generating entangled states with different frequency and temporal correlations, and the quantum advantage in terms of higher resolution these states provide, is detailed elsewhere~\cite{ravi_adhikari2025}.

 In the following, we set Alice's clock to act as the master clock, and the estimated relative clock offset is updated on Bob's clock to synchronize it with Alice's clock.  We note that others have suggested that a check on the validity of the quantum signal can also be done at this phase through the quantum bit error (QBER) metric. However, this is only a `pragmatic' check (difficult for a resource-constrained adversary to overcome); attacks on the QSTT system are not guaranteed to manifest themselves through observation of the QBER
 in the presence of a true quantum adversary (one only constrained by the laws of physics). QBER estimation does form a critical part of the QKD process. It is the inclusion of QKD that can provide information-theoretic security within QSTT.
\begin{figure}[!ht]
\centering
\resizebox{1.0\columnwidth}{!}{%
\begin{tikzpicture}[
    node distance=6mm,
    box/.style={
      rectangle, draw, rounded corners, align=center,
      minimum width=0.85\columnwidth, minimum height=1.1cm, font=\small
    },
    arrow/.style={-{Latex[length=3mm]}, thick}
]

\node[box, fill=blue!5] (quantum)
  {Quantum Signaling Channel\\(Entangled photons for timing and QKD key generation)};

\node[box, below=of quantum] (generation)
  {QKD usage rate ($r_1$) and QKD key generation rate ($r_2$)\\with $r_1 \le r_2$};

\node[box, below=of generation, fill=green!5] (infoenc)
  {Information-Theoretic Encryption\\
   $\bullet$ Encrypt subset of diff-time-tags using QKD keys (rate $r_1$)\\
   $\bullet$ Maximize tag coverage under $r_1 \le r_2$};

\node[box, below=of infoenc, fill=yellow!10] (obfuscation)
  {Obfuscated Sequence Encryption\\
   $\bullet$ Information-theoretic encryption of an instruction sequence applied 
   \\ to unencrypted diff-time-tags };

\node[box, below=of obfuscation, fill=orange!10] (output)
  {Secure Time Transfer Output\\(Encrypted and MAC-Authenticated)};

\draw[arrow] (quantum) -- (generation);
\draw[arrow] (generation) -- (infoenc);
\draw[arrow] (infoenc) -- (obfuscation);
\draw[arrow] (obfuscation) -- (output);

\end{tikzpicture}}
\vspace{4pt} 
\caption{Quantum signaling for timing and key generation. QKD-generated keys (at rate $r_2$) enable QKD-OTP encryption of a subset of diff-time-tags (provided $r_1 \le r_2$), while the remaining diff-time-tags are protected through obfuscated sequence encryption using  multiple post-quantum cryptography solutions.}
\label{fig:qstt_new_scheme}
\end{figure}
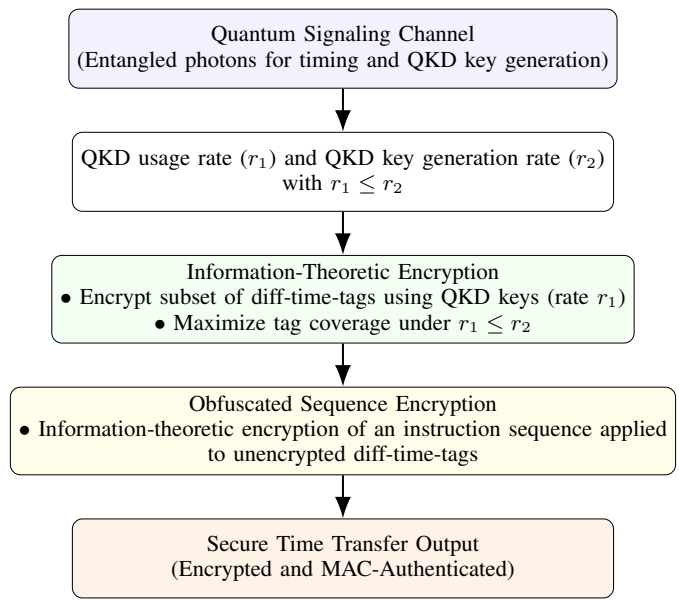

\subsection{QSTT System}\label{Subsec: QSTT}
 The main purpose of the QSTT system is to encrypt and authenticate as many time data as possible, in the most secure manner possible. A high-level overview of the system is shown in Fig.~\ref{fig:qstt_new_scheme}.
  For clarity of exposition, we break down the system into different stages, providing a general overview of each stage.
 Detailed implementation of each stage, including how the rates $r_1$ and $r_2$ are determined as an average over some timescale, is given in the supplementary material~\cite{ravi_adhikari2025}.

\subsubsection{Distribution of entangled photon pairs to users} Consider two users, Alice and Bob, who have local independent clocks that are initially unsynchronized. Let a satellite act as a mediator to distribute the entangled photon pair to users.
The users register photons locally using their detection setup. Let $O$ be a global time reference to represent the detection time-tags  of Alice and Bob. 
Let $\boldsymbol{T_A}$ and $\boldsymbol{T_B}$ denote time-tag arrays that record photon detection time-tags at Alice’s and Bob’s stations, respectively, acquired over a total measurement time $T_{run}$, the time-tags defined by:
\begin{equation}\label{equ1}
t_{\alpha} = t^o + \frac{L_{\alpha}}{c} + \delta t_{\alpha}, \quad \text{where } t_{\alpha} \in \boldsymbol{T_{\alpha}}, \; {\alpha} \in \{A, B\},
\end{equation}
 \( t^o \) is the birth time of the biphoton according to the global reference \( O \), \( L_{\alpha} \) represents the distance from the satellite to each station, \( \delta t_{\alpha} \) denotes the local clock offset relative to the global reference and \( {\alpha} \in \{A, B\} \) distinguishes Alice and Bob.

\subsubsection{Secure Timing Data Encryption} \label{subsection: Secure timing data encryption}

 In operational mode, Alice sends her encrypted  array ($\boldsymbol{T_A^*}$)  along with the message authentication code (MAC) tag over the classical channel to Bob.
 Within QSTT, we use QKD-generated secret keys (called QKD keys) to provide a way to information-theoretic secure (encrypt) as much data as possible.  
Let $\boldsymbol{T_A}=\{  t_1, t_2, \dots,  t_{n} \} $ be Alice's time-tag array of photon arrival times, with \( n \) being the number of photons detected over the measurement time \( T_{run} \). We  first transform  $\boldsymbol{T_A}$ into an array of inter-arrival times, $\boldsymbol{\Delta T_A}  = \{ \Delta t_1, \Delta t_2, \dots, \Delta t_{n-1} \} $,  where $\Delta t_j = t_{j+1} - t_j$ represents the time difference between consecutive time-tags.  We refer to $\Delta t_j$'s as diff-time-tags  and $\boldsymbol{\Delta T}$'s as   diff-time-tag arrays. We wish to QKD-OTP encrypt as many of the  $\Delta t_j$'s as possible.

 Alice cannot encrypt all diff-time-tags.\footnote{Beyond the time-tag creation rate usually exceeding $r_2$, not all of $r_1$ can be  dedicated to QKD-OTP encryption of the diff-time-tag data. Although an \textit{ a~priori} PSK of extremely large length could allow PSK-OTP encryption of all timing data (at least up till some point), storing such a large key in both users is impractical (motivating in the first place the use of QKD for scalable key generation).} Therefore, an efficient selection of which parts of the diff-time-tag array to encrypt is necessary. Alice QKD-OTP encrypts  $t_1$ (denoted by \( t_1^* \)) as it is essential to reconstruct $\boldsymbol{T_A}$ from $\boldsymbol{\Delta T_A}$.

For the diff-time-tag array, Alice first sequentially divides $\boldsymbol{\Delta T_A}$ into $2^k$ partitions, where $k$ determines the number of partitions.
 She then shuffles the partitions to break the temporal locality of the diff-time-tags. Alice records in detail the shuffle, as a set of instructions, termed $\rho$. This removes the sequential structure of the diff-time-tags, which complicates any attempts to infer the sequence.  Following this shuffling, Alice randomly selects a subset of indices,  $\mathcal{Q}$, from $\boldsymbol{\Delta T_A}$,  the subset being of size $|\mathcal{Q}| \le |\boldsymbol{\Delta T_A}|$, 
 and QKD-OTP  encrypts only the diff-time-tags corresponding to those indices. The set $\mathcal{Q}$ is recorded by Alice.
 The permutation $\rho$ QKD-OTP is encrypted and securely shared with Bob, whereas the set $\mathcal{Q}$ is sent to Bob unencrypted.  
 We note that now the diff-time-tag array can be divided into two parts: the encrypted portion, denoted by \( \boldsymbol{\Delta T_A^{\text{enc}}}  \), and the unencrypted portion, denoted by \( \boldsymbol{\Delta T_A^{\text{unenc}}}  \).
 The complete algorithmic details of this process are presented in a supplementary material~\cite{ravi_adhikari2025} alongside a discussion of the benefits and trade-offs of this specific form of permutation/encryption relative to other possibilities.

In addition to the limited QKD-OTP encryptions, we also implement an obfuscated encryption-decryption process to add a pragmatic security layer to the array of unencrypted diff-time tags ($\boldsymbol{\Delta T_A^{\text{unenc}}} $). The obfuscated encryption-decryption is defined as multiple cascaded encryption (and its corresponding decryption) of data based on an instruction sequence (IS). The IS string contains information on the order in which various encryption algorithms are to be applied.  
The IS is represented as $ (\xi_1, \xi_2, \dots, \xi_m), \text{where }
\xi_i \in \{ \text{AES}, \text{Ascon }\}   \text{~and } i = (1, 2, \dots, m)$, where $m$ is the number of times an encryption-decryption protocol is applied. The obfuscated encryption-decryption proceeds as follows: Alice derives an IS from $m'$ possible instructions,
which is used to encrypt $\boldsymbol{\Delta T_A^{\text{unenc}}}$, 
which is then PSK-OTP encrypted, producing the ciphertext. 
The encryption of the above-mentioned unencrypted diff-time tags is denoted by \( \boldsymbol{\Delta T_A^{\text{enc\_IS}} } \) which we can write as $\boldsymbol{\Delta T_A^{\text{enc\_IS}}} =\xi_m( \dots \xi_2( \xi_1(\boldsymbol{\Delta T_A ^{\text{unenc}}}))) $. We let \( \boldsymbol{T_A^*} \) denote the final encrypted timing data, comprising \( t_1^* \), \( \boldsymbol{\Delta T_A^{\text{enc}}}  \), and \( \boldsymbol{\Delta T_A^{\text{enc\_IS}}}  \), expressed as:$~\boldsymbol{T_A^*} = \left( t_1^*, \boldsymbol{\Delta T_A^{\text{enc}}} , \boldsymbol{\Delta T_A^{\text{enc\_IS}}}  \right)$. 

\subsubsection{{Authentication via information-theoretic MAC}}

Message authentication code (MAC), more specifically, Wegman-Carter  MAC is applied on the final encrypted timing data to provide information-theoretic secure authentication using QKD keys~\cite{bibak2022authentication}. The MAC tag is generated as: $\text{MAC} ( \boldsymbol{T_A^*}) = H_{k_1}(\boldsymbol{T_A^*}) \oplus k_2$, where \( H_{k_1} \) is a strongly universal hash family indexed by a fixed subset of the PSK, \( k_1 \). This hash function maps \( \boldsymbol{T_A^*} \) to an $l$-bit output tag, \( k_2 \) is a fresh subset of QKD keys (or PSK in case no QKD keys) of length \( l \) bits, generated independently for each \( \boldsymbol{T_A^*} \), and \( \oplus \) denotes bitwise XOR. Upon receiving the final encrypted timing data with MAC tag from Alice, Bob verifies its authenticity by re-computing $\text{MAC}(\boldsymbol{T_A^*}) = H_{k_1}(\boldsymbol{T_A^*}) \oplus k_2$ and checking if it matches the received MAC tag. This ensures the integrity of the transmitted time-tag array and enables the detection of any tampering.

\subsubsection{Security Considerations}

In our QSTT, Alice's array $\boldsymbol{T_A}$ is securely transmitted to Bob to estimate the relative clock offset. To synchronize Bob's clock with Alice, we estimate the relative clock offset value between the clocks, i.e., $\delta t_B-\delta t_A$. Note, each time-tag in $\boldsymbol{T_A} = \{\, t_i=t_i^0 + L_A/c + \delta t_A \mid i = 1, 2, \ldots, n \,\}$ inherently contains the local clock offset $\delta t_A$, making it vulnerable to intercept. To address this, we decoupled the local clock offset $\delta t_A$ by transforming $\boldsymbol{T_A}$ into a sequence of inter-arrival times: $\boldsymbol{\Delta T_A} = \{\, \Delta t_i = t_{i+1} - t_i = t_{i+1}^o -t_{i}^o \mid i = 1, 2, \ldots, n-1 \,\} $. The newly generated diff-time-tag array $\boldsymbol{\Delta T_A}$ contains no information about $\delta t_A$. $t_1$, which contains information about $\delta t_A$, is transmitted with QKD-OTP encryption to allow Bob to reconstruct $\boldsymbol{T_A}$. 
This ensures that only users having QKD-derived keys can synchronize their clocks with Alice's. In addition, shuffling and random diff-time-tag encryption break the temporal locality within $\boldsymbol{\Delta T_A}$ and hide as much information of the diff-time-tag as possible under the rate constraint ($r_1\le r_2)$. A discussion on the security of hybrid QKD-PQC systems under side-channel attacks can be found in~\cite{rani2025obfuscatedquantumpostquantumcryptography}.

\subsection{Clock Synchronization}\label{rob3} 

A target application we have is satellite-based QSTT. We note that in this application space it can be challenging to transfer time under the anticipated noisy and lossy channel conditions due to low coincidence detection. Providing a clear coincidence detection peak after correcting for time offsets and timing drifts is therefore critical.  We describe this process next.

On receiving Alice's final encrypted timing data, Bob first verifies the MAC tag, and if verification is successful, Bob de-encrypts (using the inverse of the encryption instructions and sequences), and uses a cross-correlation-based synchronization algorithm to estimate the relative clock offset. 
From Eq. (\ref{equ1}), we form:
$\Delta t_{BA} \equiv t_B - t_A = \frac{L_B - L_A}{c} + \delta t_{BA}$,
 where \( \delta t_{BA} = \delta t_B - \delta t_A \) is the relative clock offset. Suppose $L_A$ and $L_B$ are known exactly.\footnote{This makes our synchronization protocol a one-way transfer scheme. However,  it is straightforward to encapsulate two-way transfer schemes (including those based on non-entangled single photons) within our proposed QSTT system. These other synchronization schemes would have the same additional layers of security outlined here, at the cost of additional hardware.} Therefore, in principle, we can determine the relative clock offset $\delta t_{BA}$
 and by regularly determining this offset correct for any clock drift thereby maintaining the synchronization between Alice and Bob. To achieve this some pre-processing of the time tags are first initiated (partitioning into shorter time array blocks), followed by an initial coarse alignment and then a fine alignment. During coarse alignment, a step-size scan is performed,
making the algorithm less susceptible to processing delays caused by large time offsets. Large offsets are possible in Earth-to-satellite channels due to potential large path differentials to Alice and Bob (from the satellite). For details of this pre-processing, the reader is referred to~\cite{rani2025obfuscatedquantumpostquantumcryptography}. 

The precise clock offset between the clocks at Alice and Bob is estimated using a cross-correlation function. 
Consider the two time-tag arrays  $\boldsymbol{T_A}$ and $\boldsymbol{T_B}$, which detail the photon's arrival time recorded at Alice's and Bob's detectors, respectively, for a measurement time $T_{run}$. These time tag arrays can be translated into a distribution given by:$~A(t) = \sum_{i} \delta\bigl(t - t_A^i\bigr), \quad
    B(t) = \sum_{j} \delta\bigl(t - t_B^j\bigr)$, where $\delta$ is the Dirac function.
Mathematically, the cross-correlation function is given by:$~ C_{AB}(t') = \int A(t) \, B(t + t') \, dt$, where $t'$ is a relative delay parameter. In practice, \(C_{AB}(t')\) measures the number of coincident detection events, within a coincidence window $\Delta t$, between the two time-tag arrays $\boldsymbol{T_A}$ and $\boldsymbol{T_B}$ as a function of the relative delay \(t'\), which approaches the mathematical definition as $\Delta t\rightarrow0$. The value of $t'$ that maximizes $C_{AB}(t')$ corresponds to the relative clock offset (${\delta t_{BA}}$) between the two clocks. In our previous paper \cite{rani2025obfuscatedquantumpostquantumcryptography}, we proposed a cross-correlation-based synchronization algorithm that corrects for relative clock offset and clock drift. 
 To estimate the true relative clock offset through the $C_{AB}(t')$, it is important to ensure that the peak of the $C_{AB}(t')$ is significantly larger than random coincidences (accidental coincidences caused by noise). In our experiment, we ensure that the peak value is significantly higher than that of random coincidences.

\section{Experimental Demonstration} \label{results}
\begin{figure*}[htp] 
    \centering
    \includegraphics[width=0.95\linewidth]{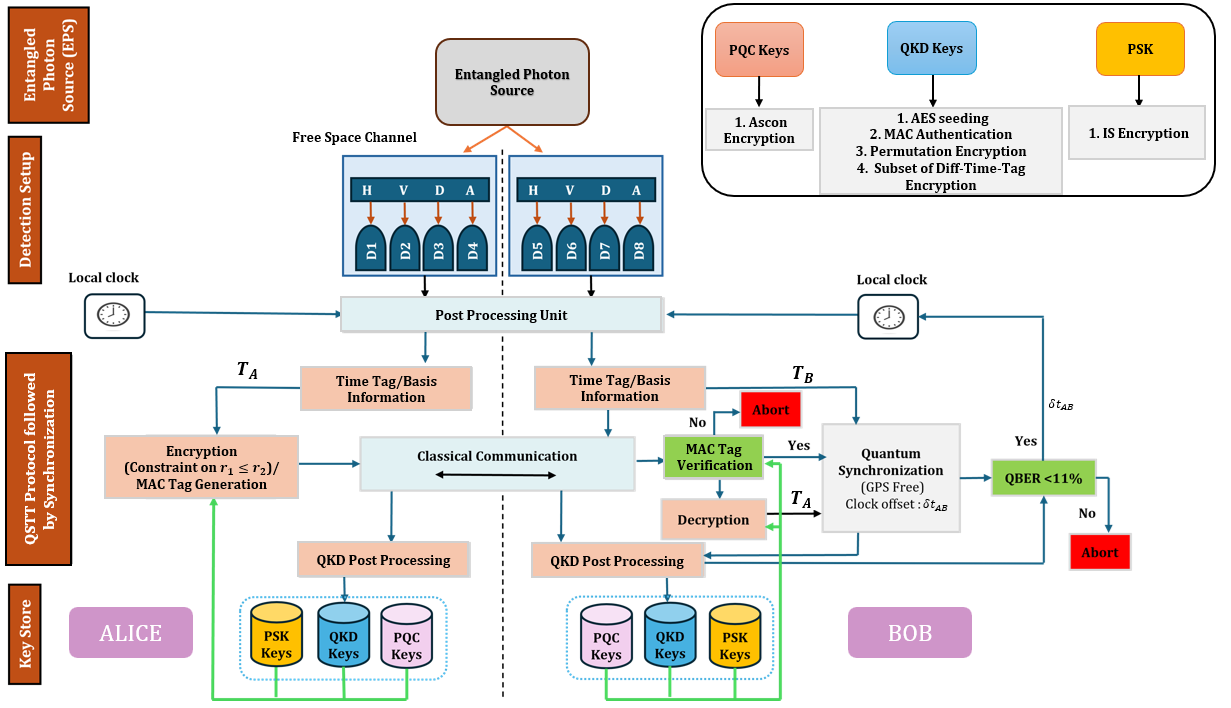}
    \vspace{4pt} 
    \caption{The experimental setup with the QSTT protocol embedded within a wider hybrid QKD-PQC system. The optical setup has single photon detectors D1 through D4 at Alice detecting the horizontal ($H$), vertical ($V$),  diagonal ($D$), or anti-diagonal ($A$) basis states, respectively (D5 through D8 similarly for Bob); and the post-processing unit, followed by encryption-decryption, MAC authentication of the final encrypted timing data, GPS free clock synchronization, generation of QKD keys. The entangled photon source is external to both Alice and Bob. In many configurations, the source is anticipated to be on board a satellite. The inset at the top right details the use of the three key sets as applied to encryption and authentication of different data (see main text).
    }
    \label{fig:setup}
\end{figure*} 

The experimental setup is shown in Fig.~\ref{fig:setup}. The QSTT for clock synchronization is embedded in a hybrid QKD-PQC system coupled to a rate constraint $r_1 \le r_2$  to form a novel QSTT system. Highly entangled photons from an entangled photon source (EPS) are distributed to two distant users, Alice and Bob, via two independent free-space quantum channels. The EPS generates hyperentangled photon pairs, known as the signal and idler, entangled in both energy-time and polarization. In the polarization degree of freedom, the state is given by $|\psi\rangle=\dfrac{1}{\sqrt{2}}(|H\rangle_{S} |V\rangle_{I} + |V\rangle_{S} |H\rangle_{I})$, where $|H\rangle$ and $|V\rangle$ represent the horizontal and vertical polarization of the photon; and the subscripts $S$ and $I$ represent the signal and idler, respectively. 

Alice and Bob receive the photons and measure the polarization of the photons projecting them onto the linear ($H/V$) and diagonal ($D/A$) bases using their four single-photon detectors, denoted $H$, $V$, $D$, and $A$. The arrival times of the detected photons are recorded in arrays 
using a time-tagging device. These arrays are continuously transmitted to Alice and Bob's personal computers for post-processing. In post-processing, Alice and Bob execute the QKD protocol, which includes transfer of classical data, QSTT, synchronization, key shifting, error estimation, error correction, and privacy amplification, resulting in the generation of QKD keys. A complete session is achieved following the privacy amplification. This QKD session takes a time $T_{run}$ to complete. The details of QKD implementation are given elsewhere~\cite{rani2025,rani2025obfuscatedquantumpostquantumcryptography}. For QSTT, Alice and Bob use the QKD keys generated between them in previous QKD sessions, as well as some use of the PSK, and execute the proposed protocol (see earlier discussions) to maintain synchronization between the two clocks. 

The experimental demonstration is conducted over a free space channel length of 1.5~m with an overall signal loss of 10.3 dB due to transmission, optical, receiving, and detection losses. The EPS source (QES2, Quibitekk) is based on Hong-Ou-Mandel interferometry, generating degenerate photon pairs (signal and idler) at 810~nm using a type II nonlinear ppKPT crystal. The generated photon pairs are inherently energy-time entangled and manifest an ultra-fast time scale in their birth-time difference. The source can be set to different power values to achieve a higher photon pair generation rate with different visibility values. In our experiment, the source is set at milliwatt power values to produce photon pairs at a rate of 10,000~$\text{s}^{-1}$, with a visibility of $87.3~\%$. A QKD detection setup used an Excelitas single-photon counting module (SPCM) with a dark count rate of 100~$\text{s}^{-1}$. The SPCM output is connected to a device (supplied by Quantum Machines) with a timing resolution of 1~ns (nanoseconds). This timestamps photon arrival times, recording them for QKD post-processing. A single QKD session took $T_{run}=4~$seconds. 

 In our protocol, we encrypted the timing information using the method described in Sec.~\ref{subsection: Secure timing data encryption}, with parameters: average number of photon detections per second ($n/T_{run}$) =~$5.5\times10^5$~$\text{s}^{-1}$, $2^k =~64$, $|Q|=~64$, $b=10$, $m=3$, $m'=4$, and MAC tag length $l$ =~61. The four possibilities of the IS are :~$00 \rightarrow (\text{ASCON, AES})$, $01 \rightarrow (\text{AES, ASCON})$, $10 \rightarrow (\text{AES, ASCON, AES})$ and~$11 \rightarrow (\text{ASCON, AES, ASCON})$. The PQC keys are established via the PQC key encapsulation mechanism~\cite{bos2018crystals}) and are used as a seed for ASCON encryption.  The roles of PQC, QKD, and PSK in timing data encryption are illustrated in the top right inset of Fig.~\ref{fig:setup}. The PQC keys are used only for Ascon encryption. The QKD keys are used for five purposes; as seeds for AES, MAC authentication, encryption of the  permutation $\rho$, $t_1$, and the subset of diff-time-tag array that are allowed to be QKD-OTP encrypted. The PSK is consumed as needed and used for the PSK-OTP encryption. Note, however, that if the QKD keys are unavailable at any time, part of the PSK may be used in their place 
 (but not for the encryption of any timing data). If the PSK is exhausted, then the QKD keys are used in their place. If both the QKD keys and the PSK are exhausted, no further generation of QKD keys is possible, and the  system reverts to the use of the PQC keys for all purposes. Importantly, each binary bit of the PSK and QKD keys is used only once.

\begin{figure}[ht]
		\centering		
        \includegraphics[width=0.85\linewidth]{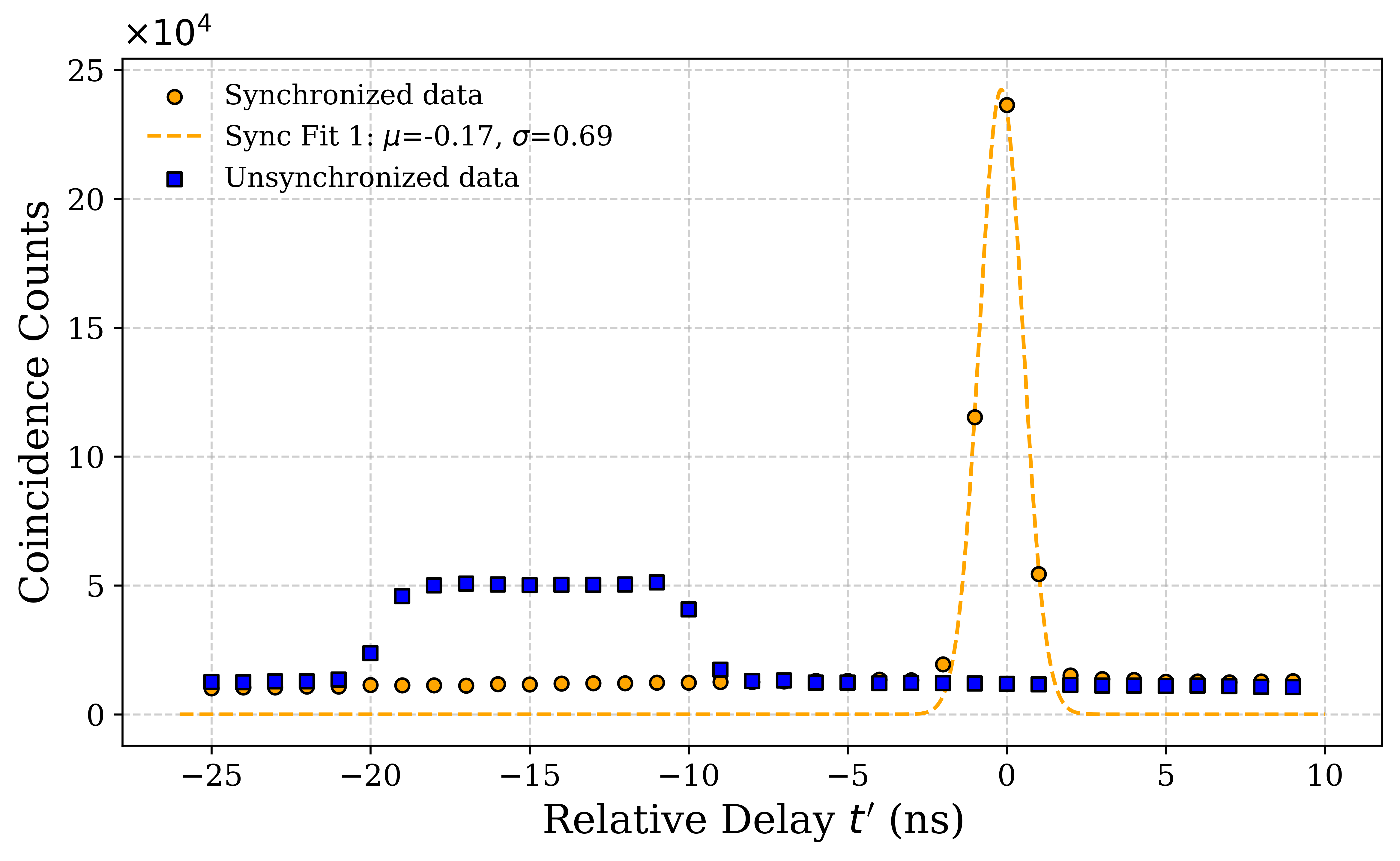}
		\caption{Coincidence counts as a function of a relative delay, $t'$, added to Bob’s time-tag array. Counts are shown with (orange) and without (blue)
application of the synchronization protocol. For each $t'$, coincidences are counted only when Alice’s and Bob’s time-tags match within a 0.5~ns window. The data is accumulated over $40~$seconds.} 
        \label{fig:delay_comparison_plot}
\end{figure}

We used the synchronization algorithm described in~\cite{rani2025obfuscatedquantumpostquantumcryptography}, testing it within this new setup by introducing a 10~ns relative clock offset and a clock drift of 0.25~ns/sec in Bob's time-tag array, which transformed $\boldsymbol{T_B}$ to $\boldsymbol{T_B'}$. Bob executed the synchronization algorithm on $\boldsymbol{T_A}$ and $\boldsymbol{T_B'}$,
and estimated the relative clock offset. 
Bob then updated his time tag array $\boldsymbol{T_B'}$ according to the estimated relative clock offset. The continuous execution of the synchronization algorithm also eliminated the clock drift.  
To verify synchronization performance, we computed the cross-correlation function (see Sect.~\ref{rob3}) for synchronized and unsynchronized datasets, accumulated over 40~seconds. As shown in Fig.~\ref{fig:delay_comparison_plot}, the synchronized data exhibited a sharp coincidence peak at zero relative delay, confirming the complete removal of the relative clock offset and associated synchronization error. In contrast, the unsynchronized data exhibited a broad distribution with lower coincidence counts spread over a relative delay range of approximately -10~to -20~ns, consistent with the introduced relative clock offset and clock drift. Therefore, we have  retrieved a sharp coincidence peak at zero relative delay.

Fig.~\ref{fig:delay_comparison_plot} also illustrates the precision of our synchronization process. The imperfect synchronization is a consequence of the timing jitter of the subsystems linked to the system, such as detectors, time-taggers, clocks, and the inherent jitter of the biphoton source. The standard deviation of the timing jitter can be written as: $~\sigma= \sqrt{\sigma_{int}^2 + 2\sigma_{det}^2 + 2\sigma_{t.t}^2 },$ where $\sigma_{int}$ is the intrinsic temporal width of the biphoton, and $\sigma_{det}$ and $ \sigma_{t.t}$ are the timing jitter of the detector and the time-tagger unit, respectively. Instead of directly measuring the timing jitter of these individual subsystems, we instead estimate $\sigma$ by modeling the peak distribution of Fig.~\ref{fig:delay_comparison_plot} as a Gaussian function. By this process, we estimate the standard deviation of our synchronization to be 0.69~ns.

\begin{figure}[ht]
		\centering		
        \includegraphics[width=0.9\linewidth]{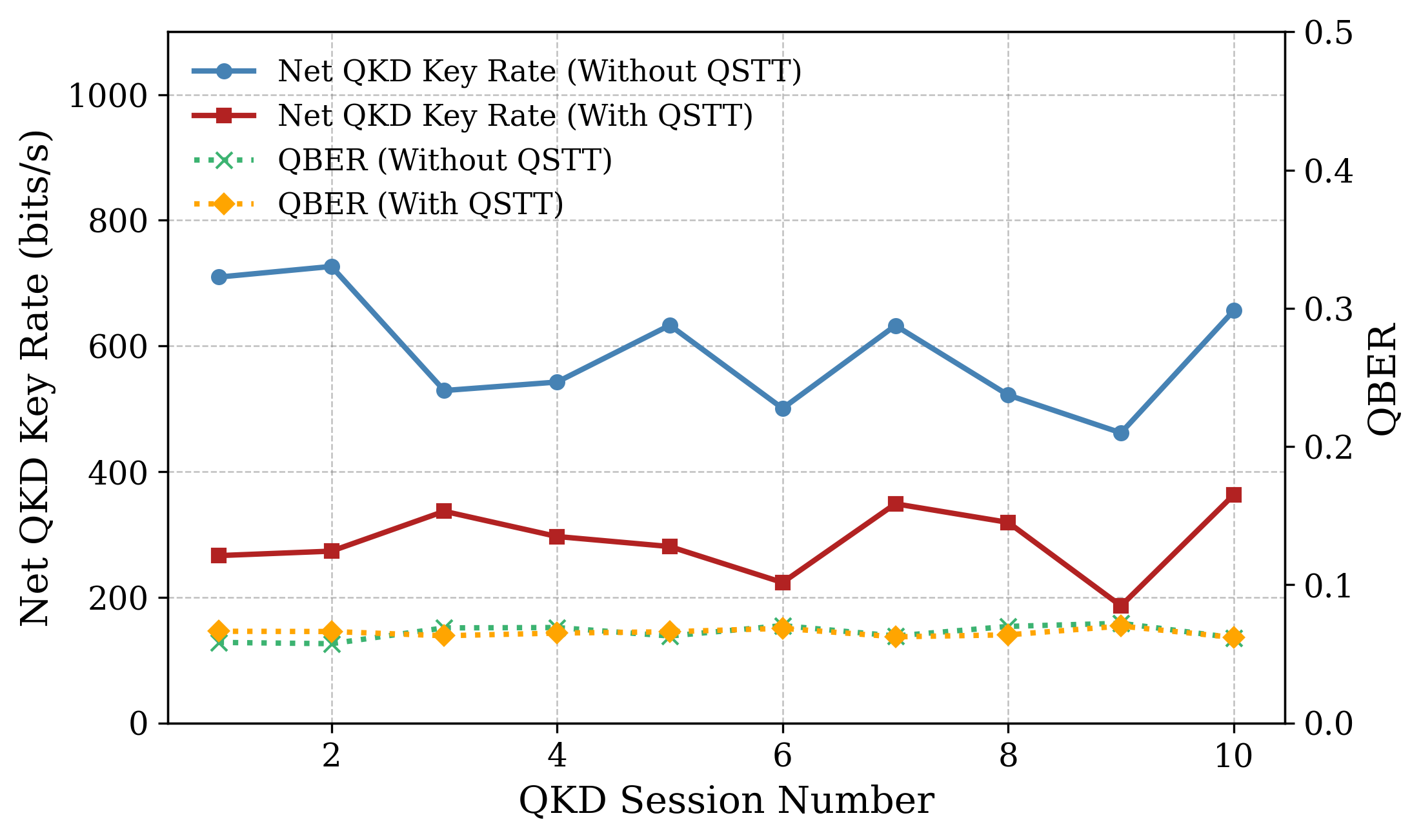}
		\caption{Net QKD key rate and QBER across $10$ QKD sessions with and without the QSTT protocol. The QKD key rate decreases under a secure synchronization protocol due to key consumption in QSTT, while the QBER remains stable. This trade-off highlights the balance between synchronization security and key generation efficiency in QKD systems. Since the key rates shown assume the asymptotic limit of data sampling, they are upper limits.}
        \label{fig:keyrate_qber_vs_sessions}
\end{figure}

In Fig. \ref{fig:keyrate_qber_vs_sessions}, we show a net QKD key rate (generated QKD key rate - consumed QKD key rate) and QBER over $10$ QKD sessions with and without our proposed protocol. The lower net QKD key rate on the use of our protocol arises from the consumption of QKD keys from the previous session during QSTT, including the encryption of specific QKD-OTP data (mentioned in Sect. \ref{subsection: Secure timing data encryption}), the seeding of AES-256, and MAC authentication.
On average, with our proposed QSTT protocol, the QKD system effectively generated $289\pm56$ bits/s after the consumption of QKD keys in QSTT. In addition, we note that the QBER remained within acceptable limits for QKD key generation across all sessions.
Hence, we can see that the integration of our QSTT protocol within a hybrid QKD-PQC system achieved a useful net QKD key rate. For comparison, the net QKD key rate of a QKD-PQC prototype system without applying our proposed QSTT system was $664 \pm 102$~bits/s. This approximate factor of two difference in key rates demonstrates the impact of the additional security brought about by the integration of the QSTT system.

\section{Conclusion} \label{Sec:conclusion}

We designed and experimentally demonstrated a new QSTT protocol, combined within a hybrid QKD-PQC architecture. In our system, entangled photons acted as a timing signal and a source for QKD key generation. 
Operating under a rate constraint, the generated QKD keys were used to ensure information-theoretic security on the maximum amount of diff-time-tag data. As an additional security layer, a hybrid QKD-PQC instruction sequence  was information-theoretically encoded with a pre-shared key, and a permutation ($\rho$) was used to scramble the timing data into a non-sequential temporal array, with $\rho$ QKD-OTP encrypted.  We believe that the new QSTT system reported here represents the most robust quantum secure time transfer system implemented to date. Its use to secure the synchronization of satellite-based communication networks should be of particular value.

\section*{\label{sec:Acknowledgments}Acknowledgments}
The authors thank Dr. Dushy Tissainayagam from Northrop Grumman Australia (NGA) for useful discussions.
This research has been carried out as a project co-funded by NGA and the Defense Trailblazer Program, a collaborative partnership between the University of Adelaide and the University of New South Wales, co-funded by the Australian Government, Department of Education.

{\small
\bibliographystyle{IEEEtran}
\bibliography{IEEEabrv,ref}
}
\clearpage
\begin{center}
    \vspace*{0.5cm}
    {\LARGE {A New Quantum Secure Time Transfer System: Supplementary Material}}\\[1em]

\end{center}
\maketitle
\thispagestyle{empty}
\pagestyle{empty}

This supplementary material presents additional details supporting the manuscript ``A New Quantum Secure Time Transfer System,' by Ravi Singh Adhikari, Aman Gupta, Anju Rani, Xiaoyu Ai, and Robert Malaney.

\section{Algorithmic Details}

\begin{algorithm}
\caption{Data Encryption of Timing Data}\label{encryption_algo}
\small
\SetAlgoLined
\DontPrintSemicolon
\SetKwComment{Comment}{/* }{ */}

\KwIn{the time-tag array $\boldsymbol{T_A}$;  bits available from previous QKD session $r_2 T_{\text{run}}$; bit length $b$; the measurement time $T_{\text{run}}$; number of partitions $2^k$; bits consumed for MAC authentication $k_2$; bits consumed for AES seeding $k_{\text{AES}}$.}

\KwOut{Final encrypted timing data $\boldsymbol{T_A^*}$; encrypted permutation $\rho^*$; random subset of indices $\mathcal{Q}$; encrypted instruction sequence $\pi$.}
\BlankLine  

$\boldsymbol{\Delta T_A} \gets \text{ConvertToInterArrivalTime}(\boldsymbol{T_A})$ \\
\CommentSty{/* We refer to $\boldsymbol{\Delta T_A}$ as the diff-time-tag array */}
\BlankLine  

$t_1^* \gets \text{QKD-OTP\_Encrypt}(t_1)$\\ \CommentSty{/* Encrypt $t_1$ using QKD-OTP; consumes $k_{t_1}$ key bits */}
\BlankLine  

$\mathcal{S} \gets \text{Partition}(\boldsymbol{\Delta T_A}, 2^k)$\\ \CommentSty{/* Divide $\boldsymbol{\Delta T_A}$ into $2^k$ partitions */}\\
\BlankLine  
$\rho \gets \text{Permutation}(\mathcal{S})$ \\\CommentSty{/* Derive permutation to shuffle partitions */}\\
\BlankLine  
$\rho^* \gets \text{QKD-OTP\_Encrypt}(\rho)$ \\\CommentSty{/* Encrypt permutation $\rho$ and share with Bob */}\\
\BlankLine  
$\mathcal{S}^{\rho} \gets [S_{\rho(1)}, S_{\rho(2)}, \ldots, S_{\rho(2^k)}]$\\ \CommentSty{/* Shuffle partitions as per $\rho$ */}
\BlankLine  

$n-1 \gets \text{length}(\boldsymbol{\Delta T_A})$ 

\BlankLine  

$|\mathcal{Q}| \gets \left\lfloor \dfrac{r_2 T_{\text{run}} - k \cdot 2^k - k_2 - k_{\text{AES}} - k_{t_1}}{b} \right\rfloor$\\ \CommentSty{/* Maximum number of diff-time-tags that can be encrypted */}
\BlankLine  
$\mathcal{Q} \gets \text{RandomSubset}(\{1, 2, \dots, n-1\}, |\mathcal{Q}|)$ \\\CommentSty{/* Select $|\mathcal{Q}|$ random indices */}\\
\BlankLine  
$\boldsymbol{\Delta T_A^{\text{enc}}} \gets \text{QKD-OTP\_EncryptSubset}(\mathcal{S}^{\rho}, \mathcal{Q})$ \\\CommentSty{/* Encrypt the diff-time-tags corresponding to the indices contained in Q using QKD-OTP */}\\
\BlankLine  

$\text{IS} \gets (\xi_1, \xi_2, \dots, \xi_m) \gets \text{InstructionSequence}(\text{AES}, \text{Ascon})$\\ \CommentSty{/* Derive encryption sequence */}\\\BlankLine  
$\pi \gets \text{PSK-OTP\_Encrypt}(\text{IS})$\\ \CommentSty{/* Encrypt IS using PSK-OTP */}\\\BlankLine  
$\boldsymbol{\Delta T_A^{\text{enc\_IS}}} \gets \xi_m \dots \xi_2(\xi_1(\boldsymbol{\Delta T_A^{\text{unenc}}}))$ \\\CommentSty{/* Apply encryption sequence on the unencrypted diff-time-tags, $\boldsymbol{\Delta T_A^{\text{unenc}}}$ $\boldsymbol{}$ */\\}\BlankLine  

$\boldsymbol{T_A^*} \gets \left(t_1^*, \boldsymbol{\Delta T_A^{\text{enc}}}, \boldsymbol{\Delta T_A^{\text{enc\_IS}}}\right)$ \\\CommentSty{/* Construct final encrypted timing data */}\BlankLine  

\Return $\boldsymbol{T_A^*}$, $\rho^*$, $\mathcal{Q}$, $\pi$
\end{algorithm}

\begin{figure}[ht]
		\centering	
        
        \includegraphics[width=1\linewidth]{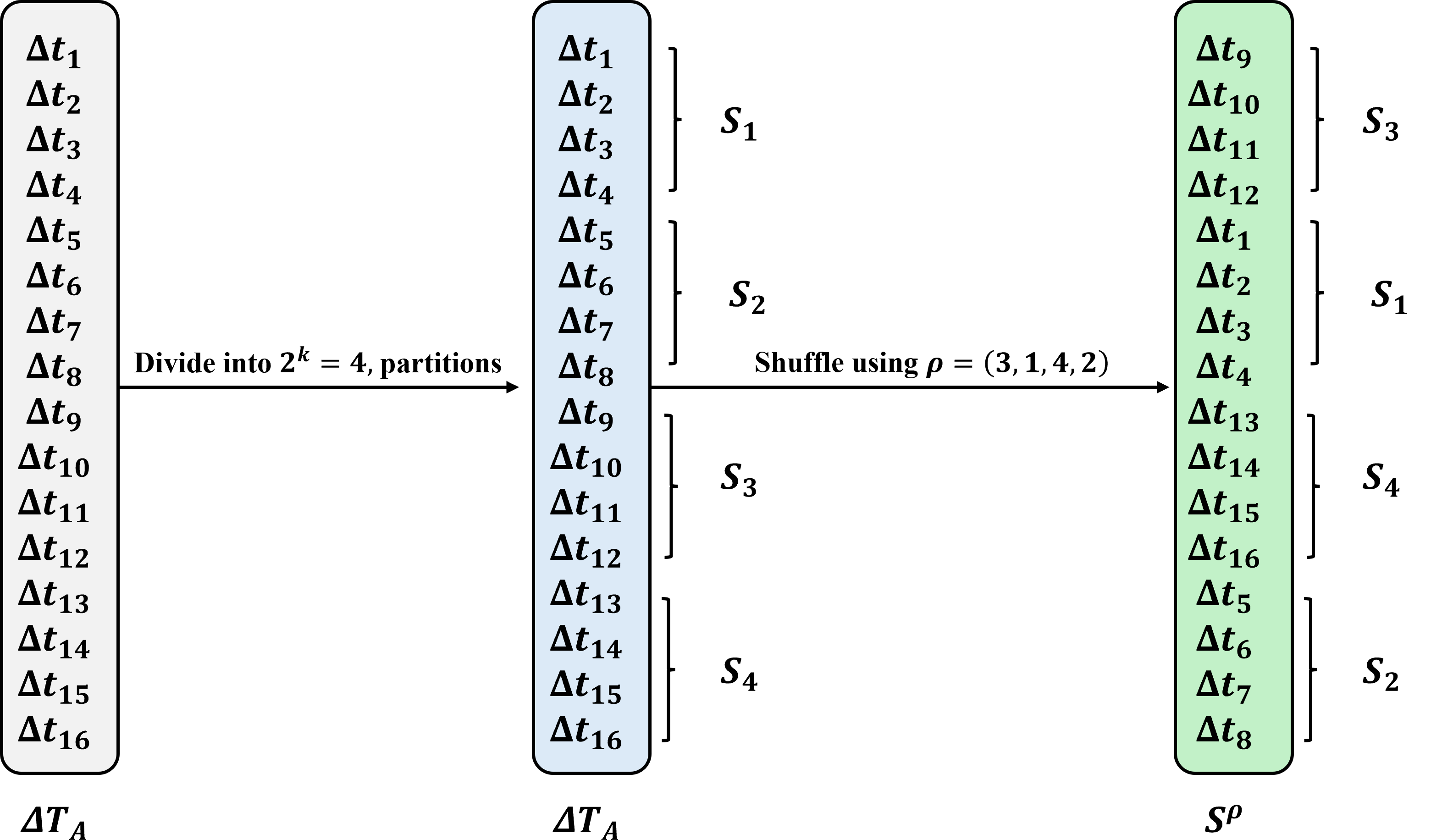}
        \vspace{4pt} 
		\caption{Example illustration of the partitioning and shuffling process for one segment of $\boldsymbol{\Delta T_A}$. For this example, $n=16$ samples are divided into $2^{k}=4$ partitions (each of size 4), and then the partitions are permuted according to $\rho=(3,1,4,2)$. The left column shows the $\boldsymbol{\Delta T_A=}\{\Delta t_{1},\dots,\Delta t_{16}\}$, the centre column shows the four contiguous partitions before shuffling, and the right column shows the final reordered sequence after applying the permutation $\rho$.
}
\label{fig: shuffling}
\end{figure}

Algorithm~1  details encryption of timing data in terms of pseudo-code.
   The design of this encryption process ensures security within the constraint imposed by the available QKD keys ($r_1\le r_2)$. While complete shuffling of the diff-time-tags might seem ideal from a theoretical standpoint (maximizing randomness), it would be at the cost of additional QKD key resources. 
This is because encrypting the  permutation $\rho$ would require a bit string of length $k'(n-1)$, where $n-1$ is the number of elements in $\rho$, which in this case is equal to the number of elements in $\boldsymbol{\Delta T_A}$,
and $k'$ is the number of bits needed to encrypt a single index in $\rho$.
To achieve a more resource-efficient approach, we instead partition the diff-time-tag array into $2^{k}$ partitions and restrict shuffling to the partition level. 
Encryption of $\rho$ then requires only $k\cdot 2^{k}$ bits ($k < k' \quad \text{and} \quad 2^k < n - 1$), where $2^{k}$ is the number of elements in $\rho$, and $k$ is the number of bits needed to encrypt a single index in $\rho$, dramatically reducing the use of keys. This partition-based shuffling maintains strong scrambling properties while ensuring that the protocol remains efficient and secure under realistic resource constraints.
 Fig.~\ref{fig: shuffling} provides a schematic view of the partitioning and shuffling process described in the algorithm through a simple example. 
   
In the main text,  $r_1$ is the QKD key
usage rate and $r_2$ the QKD key creation rate. In our implementation, the former is determined via
\begin{equation}\label{eq: key_rate}
    r_1 = \frac{k \cdot 2^k + b \cdot |Q| + k_2 + k_{\text{AES}} + k_{t_1}}{T_{run}}\le r_2,
\end{equation}
where \( k \cdot 2^k \) is the number of bits consumed in encrypting  \( \rho \), \( b \cdot |Q| \) is the number of bits consumed to encrypt a subset of the diff-time-tags, \( k_2 \) is the number of bits consumed for MAC authentication, \( k_{\text{AES}} \) is the number of bits consumed for AES seeding,  \( k_{t_1} \) is the number of bits consumed to encrypt the timestamp \( t_1 \), and $T_{run}$ denotes the measurement time of the QKD session. The QKD key creation rate $r_2$, is defined as the number of QKD bits extracted per QKD session divided by $T_{run}$. Eq.~(\ref{eq: key_rate}) also implies that the maximum number of diff-time-tags that can be encrypted is given by : 
\[
|Q| \le \left\lfloor \frac{r_2 T_{\text{run}} - k \cdot 2^k - k_2 - k_{\text{AES}} - k_{t_1}}{b} \right\rfloor,
\]
where \( \lfloor \cdot \rfloor \) denotes the floor operator that ensures that \( |Q| \) is an integer and the numerator denotes the total number of bits remaining from the previous QKD session after being utilized in the aforementioned  operations (for example, AES seeding). In our implementation, the value of \( k \) is chosen to be 6, leading to \( 2^k = 64 \) partitions and a total of \( 64! \) possible permutations $\rho$, making it computationally challenging for an adversary to determine the \( \boldsymbol{\Delta T_A} \).

It is possible to determine  $r_1$ and $r_2$ differently from the above. For example, if a finite key analysis were to be used, it would perhaps be more useful to use the time to achieve a required block length (needed for a particular security setting) as the timescale over which the averaging should occur. Regardless of the definition of these rates, the conceptual framework shown in Fig.~1 of the main text remains the same. 
Again, we caution that the key rates shown in Fig.~4 of the main text are based on an asymptotic key rate analysis and therefore represent only upper limits on the true finite key rates for some security parameter (failure probability).

\section{Timing advantage from entangled states}
Entangled photon pairs are generated through the spontaneous parametric down-conversion (SPDC) process. When a high-energy pump photon is incident on a nonlinear crystal, there is a small probability that it will be annihilated, and two lower-energy photons (signal and idler) are created. The quantum state of the photon pair generated by the SPDC process, under the assumption of a monochromatic pump, can be written in the frequency domain as~\cite{PhysRevA.31.2409,PhysRevA.56.1627,PhysRevA.80.033814,kim2002generation,linares2023characterization}
\begin{equation}\label{equ:state_frequency_domain}
    \lvert \Psi \rangle_{si} = \mathcal{N} \iint d\omega_s d\omega_i \, 
    \Lambda(\omega_s, \omega_i) \lvert \omega_s \rangle_s \lvert \omega_i \rangle_i,
\end{equation}
where $\lvert \omega_j \rangle_j$ $(j = s,i)$ denotes a single-photon Fock state at frequency $\omega_j$ for the signal~(\textit{s}) and idler~(\textit{i}) photons, respectively. Here, $\mathcal{N}$ is a normalization constant incorporating the nonlinear coefficient $\chi^{(2)}$,  the crystal length $L$, and other geometric factors. 
The function $\Lambda(\omega_s, \omega_i)$ is the joint spectral amplitude (JSA) 
of the signal and idler field, describing the amplitude of generating the two photons at the  frequencies $\omega_s$ and $\omega_i$, respectively. The JSA  is the product of the phase-matching function  and the
spectral amplitude of the pump field.
Given our assumption of a monochromatic pump, it  can be written in terms of the phase matching function $\mathbf{f}(\omega_s,\omega_i)$ as
\begin{equation}\label{equ:JSA_alpha}
    \Lambda(\omega_s, \omega_i) =
    \mathbf{f}(\omega_s,\omega_i)\, \delta(\omega_p - \omega_s - \omega_i),
\end{equation}
where  $\mathbf{f}(\omega_s,\omega_i)=\mathrm{sinc}\!\left( \frac{\Delta k L}{2} \right)$, $\Delta k = k_p(\omega_s + \omega_i) - k_s(\omega_s) - k_i(\omega_i)$ is the phase mismatch with $k_j(\omega_j) = n_j(\omega_j)\omega_j / c$ and $n_j(\omega_j)$ the refractive index experienced by the field $j$. Note that here the bracketed $(\omega_j)$ is used simply to expressly point out the frequency dependence of the parameter.

We investigate a second-order correlation function, $G_{si}^{(2)}(\tau)$, for the signal  and idler  photon pair, which can be expressed as \cite{muller2020general}: 
\begin{equation}
G^{(2)}_{si}(\tau) = 
_{si}\langle \Psi | 
\hat{E}^{(-)}_i(t)
\hat{E}^{(-)}_s(t+\tau)
\hat{E}^{(+)}_s(t+\tau)
\hat{E}^{(+)}_i(t)
| \Psi \rangle_{si},
\end{equation}
\begin{equation}\label{eq:G2}
 = | 
\hat{E}^{(+)}_s(t+\tau)
\hat{E}^{(+)}_i(t)
| \Psi \rangle_{si}|^2,
\end{equation}
where \( \hat{E}^{(\pm)} \) are the positive- and negative-frequency components of the electric field operators.
The electric field operator can be written as:
\begin{equation} \label{eq:electric field}
    \hat{E}^{(+)}_{s/i}(t) \approx \int_{-\infty}^{\infty} d\omega \, 
\hat{a}_{s/i}(\omega) \, e^{-i(\omega t - k z)},
\end{equation}
where \(\hat{a}_{s/i}(\omega)\) is the annihilation operator for the measured photon 
in the signal  or idler  mode at frequency \(\omega\).
By inserting Eqs.~(\ref{equ:state_frequency_domain}) and (\ref{eq:electric field}) into the definition of 
\( G^{(2)}_{si}(\tau) \) [Eq.~(\ref{eq:G2})], we have \cite{muller2020general}:
\begin{equation}\label{eq: solved G2}
G^{(2)}_{si}(\tau) =
\left|
\int_{-\infty}^{\infty} 
d\omega_s \, e^{-i\omega_s\tau}
\, \mathbf{f}(\omega_s, \omega_i=\omega_p - \omega_s)
\right|^2.
\end{equation}
 
Eq.~(\ref{eq: solved G2}) shows how the second-order correlation function 
$G^{(2)}_{si}(\tau)$ is related to the Fourier transform of $\mathbf{f}(\omega_s, \omega_i)$.

From this analysis it is clear that the second-order correlation function will possess a complex relation with the characteristics of the pump beam, and several physical characteristics of the crystal implementing the non-linear reaction. More detailed analysis of the interplay between these characteristics and the impact they have on the temporal correlation of the signal and idler is provided in many publications (e.g. \cite{muller2020general,timelens}). This further analysis depends heavily on various assumptions adopted; however, typically the standard deviation on the difference in signal-idler arrival times is of the order of 100-300~fs for most SPDC configurations~\cite{muller2020general}.

 In our experimental setup, we used an entangled photon pair source (QES 2.4, Qubitekk), generating degenerate photon pairs (signal and idler) at 810~nm, with a spectral bandwidth of approximately~$2$~nm (FWHM), using a type II nonlinear ppKPT crystal. Detailed experimental validation of these characteristics is reported elsewhere~\cite{otherreport}. If our photon detector had high enough resolution we would have anticipated a standard deviation on the difference of arrival times of the order of 300~fs.
 However, due to the limited temporal resolution of our current single-photon detectors and time-tagging electronics, such ultrafast timings cannot be directly resolved in our measurements. In our experiment, the tight correlation between the signal-idler is convolved with our limited hardware resolution, and we observed a standard deviation in the difference in arrival times of $\sigma = 700~\mathrm{ps}$ (see the main text). Nevertheless, advances in detection and timing technology are expected to enable reliable access to this intrinsic temporal resolution in the sub-picosecond range in the not-to-distant future. As we discuss below, this will enable an additional layer of pragmatic security to the synchronization process.

We should point out that, in principle, frequency entanglement across $M>2$  modes is possible, which can improve the time correlations mentioned above. In a series of works \cite{giovannetti2001quantum,PhysRevA.65.022309,giovannetti2002generating} an $\sqrt{M}$ enhancement is observed in the scaling performance of the time correlations for some  higher-mode states under specific geometries of the time detectors relative to the entangled source. One of these higher-mode states is a three-mode down-converted state \cite{giovannetti2001quantum} of the form
\begin{equation}\label{eq:higherMode}
    |\Psi\rangle_{\text{en}} \equiv\int d\omega d\omega' \, f(\omega, \omega') \lvert \omega \rangle \lvert \omega' \rangle |\omega_p - \omega - \omega'\rangle,
\end{equation}
 where $\omega_p$ is the pump frequency, and $\omega$, $\omega^{'}$, and  $\omega_p - \omega - \omega'$ are the frequencies of the photons generated. Here $f(\omega,\omega')$ is the JSA of the three-mode down-converted state. 
Another possibility is a state of the form
\begin{equation}\label{eq: correlted state}
    |\Psi\rangle_{\text{en}} \equiv \int d\omega \, \phi_{\omega} |\omega\rangle_1 |\omega\rangle_2 \cdots |\omega\rangle_M,
\end{equation}
where $\phi_{\omega}$ denotes the JSA of a state in which all photons share the same frequency~$\omega$.

In the configuration of Eq. (\ref{eq: correlted state}), the enhanced correlation is not in the time difference but in the average detection-time of the detectors. The difficulty in creating the higher modes states of Eqs.~(\ref{eq:higherMode}) and~(\ref{eq: correlted state}) is discussed in \cite{giovannetti2001quantum} and \cite{PhysRevA.65.022309}, respectively. We are unaware of any experimental implementation that has demonstrated this $\sqrt{M}$ enhancement in timing correlations.

We should caution that the enhancement in time correlations of frequency entangled states (relative to unentangled states of the same  mode number) does not lead to an advantage over a quantum adversary across all possible attacks. For example, a quantum adversary can carry out an interaction with a photon on a time scale approaching zero as her energy resources approach infinity (we ignore in this statement the general-relativistic implications on timings of an energy approaching infinity~\cite{malaney2}. Nevertheless, the enhanced-timing correlations that we have discussed can offer pragmatic security enhancements to a real-world adversary and are, therefore, worth pursuing.

\end{document}